\documentclass[preprint,superscriptaddress,aps]{revtex4}

\usepackage{color}
\usepackage{epsfig}

\usepackage{graphicx}
\newcommand{\kbf}      {\textbf{k}}
\newcommand{\qbf}      {\textbf{q}}

\newcommand{\Ham}      {\mathrm{H}}

\begin{document}

\title{Emergence of interfacial polarons from electron-phonon coupling in graphene/h-BN van der Waals heterostructures}

\author{Chaoyu Chen$^{1}$, Jos\'e Avila$^{1}$, Shuopei Wang$^{2}$, Yao Wang$^{3,4,5}$, Marcin Mucha-Kruczy\'nski$^{6,7}$, Cheng Shen$^{2}$, Rong Yang$^{2}$, Benjamin Nosarzewski$^{3,4}$, Thomas P. Devereaux$^{4,8}$, Guangyu Zhang$^{2}$ \& Maria C. Asensio$^{*}$}

\affiliation{ANTARES Beamline, Synchrotron SOLEIL $\&$ Universit$\acute{e}$ Paris-Saclay, L'Orme des Merisiers, Saint Aubin-BP 48, 91192 Gif sur Yvette Cedex, France.\\
$^{2}$ Beijing National Laboratory for Condensed Matter Physics and Institute of Physics, Chinese Academy of Sciences, Beijing 100190, China\\
$^{3}$ Department of Applied Physics, Stanford University, California 94305, USA\\
$^{4}$ Stanford Institute for Materials and Energy Sciences, SLAC National Laboratory and Stanford University, Menlo Park, California 94025, USA\\
$^{5}$ Department of Physics, Harvard University, Cambridge, Massachussetts, 02138, USA\\
$^{6}$ Department of Physics, University of Bath, Claverton Down, Bath BA2 7AY, United Kingdom \\
$^{7}$ Centre for Nanoscience and Nanotechnology, University of Bath, Claverton Down, Bath BA2 7AY, United Kingdom \\
$^{8}$ Geballe Laboratory for Advanced Materials, Stanford University, California 94305, USA \\
$^{*}$ maria-carmen.asensio@synchrotron-soleil.fr
}

\begin{abstract}
\textbf{
Van der Waals heterostructures---vertical stacks of layered materials---offer new opportunities for novel quantum phenomena which are absent in their constituent components.
Here we report the emergence of polaron quasiparticles at the interface of graphene/hexagonal boron nitride (h-BN) heterostructures. Using nanospot angle-resolved photoemission spectroscopy, we observe zone-corner replicas of h-BN valence band maxima, with energy spacing coincident with the highest phonon energy of the heterostructure---an indication of Fr\"ohlich polaron formation due to forward-scattering electron-phonon coupling. Parabolic fitting of  the h-BN bands yields an effective mass enhancement of $\sim$ 2.3, suggesting an intermediate coupling strength. Our theoretical simulations based on Migdal-Eliashberg theory corroborate the experimental results, allowing the extraction of microscopic physical parameters.
Moreover, renormalisation of graphene $\pi$ band is observed due to the hybridisation with the h-BN band.
Our work generalises the polaron study from transition metal oxides to Van der Waals heterostructures with higher material flexibility,  highlighting interlayer coupling as an extra degree of freedom to explore emergent phenomena.
}

\end{abstract}

\maketitle
Van der Waals heterostructures formed by the superposition of two-dimensional (2D) atomic materials have been revealed as a promising field in condensed matter physics\cite{Geim1}. 
One of the most interesting heterostructures consists of hexagonal boron nitride (h-BN) and monolayer graphene. Their small lattice mismatch ($\sim$ 1.8 $\%$) and high interfacial quality have already led to the observation of a number of exotic phenomena such as Hofstadter's butterfly\cite{Geim2, Ashoori, Kim}, fractal quantum Hall effect\cite{Kim} and commensurate-incommensurate transition\cite{Geim3}.
Like for most other crystalline materials, it is the intrinsic electronic structure (energy $\it {vs} $ momentum relation) that determines their macroscopic behaviours. From the quantum mechanical point of view, electronic bands arise from the overlap of electronic states of atoms in the lattice. Consequently, the electronic structure contains complex information about the interplay of charge, orbital, spin and lattice degrees of freedom. 
For this reason the electronic structure measurement plays an essential role in understanding the underlying physics of van der Waals heterostructures.

To this point, few techniques to measure the electronic structure of graphene/h-BN vertical heterostructures (G/h-BN) have been employed\cite{Zhou, Wang, Molen}.
Of particular importance is angle-resolved photoemission spectroscopy (ARPES) which measures directly the single-particle spectral function. ARPES studies have revealed the electronic signature of van der Waals interaction in G/h-BN\cite{Zhou, Wang}. The observed Dirac cone replicas and mini gaps are direct consequences of van der Waals potential modulation imposed on graphene. However, surprisingly little is known so far about the modulated electronic structure of h-BN in G/h-BN. Even though the unique electronic, optical and mechanical properties of graphene in G/h-BN have been well studied, the degree of disruption of the h-BN band structure at the heterostructure is still an open question. 

\begin{figure*}[tbp]
\centering
\includegraphics[width=1\textwidth]{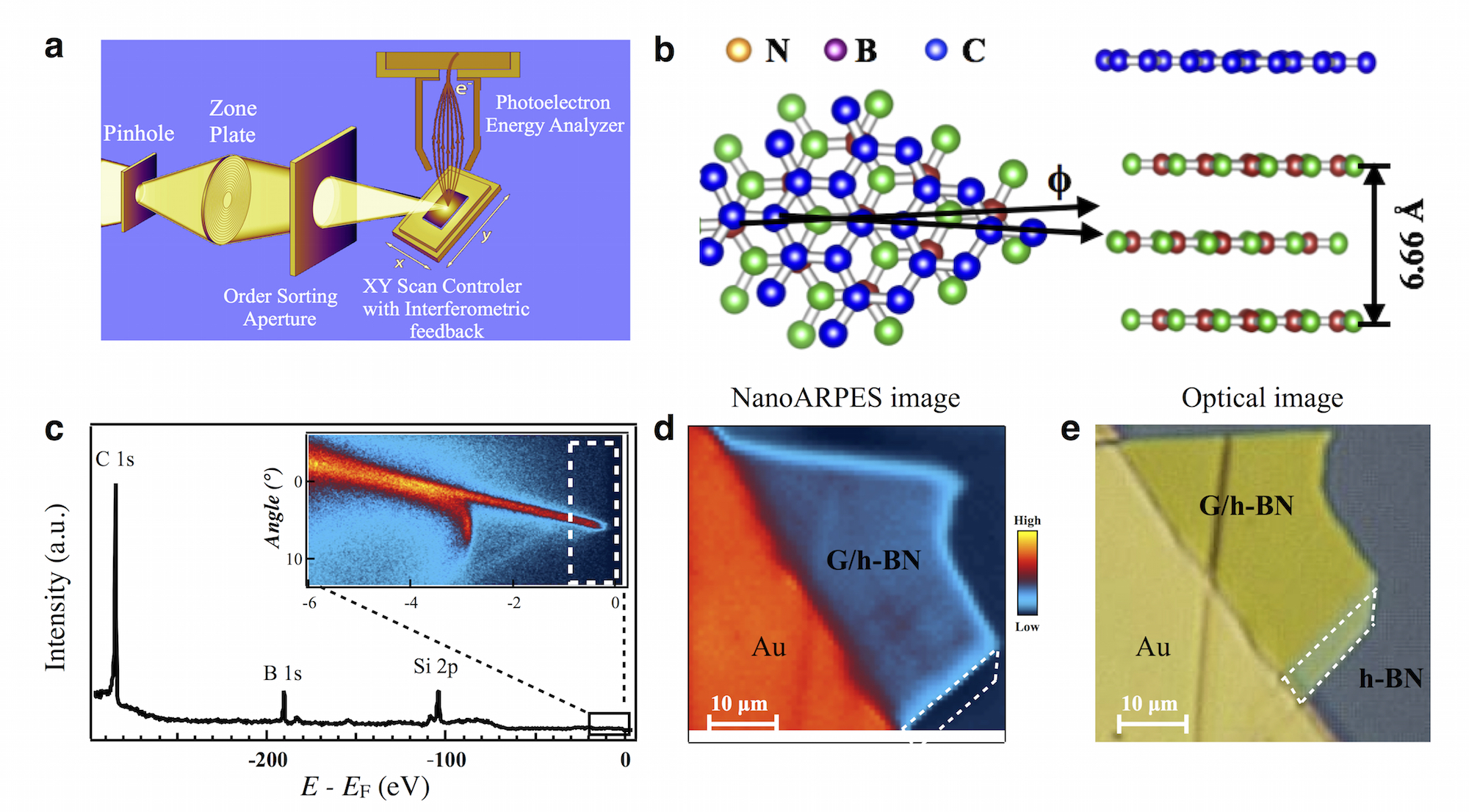}
\caption{\textbf{General experimental geometry and sample characterisation}
\textbf{a}, Schematic focus mechanism of NanoARPES. 
\textbf{b}, Schematic of G/h-BN lattice structure.
\textbf{c}, Overall photoemission core level spectrum taken with bean energy 350 eV and beam spot larger than 50 $\mu$m. The inset shows the detailed NanoARPES spectrum close to Fermi level taken with beam energy 100 eV. 
\textbf{d}, NanoARPES image shows the distribution of photoemission intensity integrated from the energy-angle window (dashed box) shown in \textbf{c}.
\textbf{e}, Optical image of the measured sample. The heterostructure is grounded by Au contact. Pristine h-BN border is highlighted by dashed white lines, in panels \textbf{d} and \textbf{e} .
}
\end{figure*}

In this study, we focus on the modulated electronic structure of h-BN in G/h-BN. Currently, high-quality G/h-BN heterostructures, especially those with high device mobility\cite{Geim2, Ashoori, Kim, Geim3}, are mainly fabricated by transfer method\cite{Dean1, LeRoy1}, with typical domains in micrometric or nanometric scale. This makes traditional ARPES (spot size $\sim$100 $\mu$m) measurements  impossible.
To measure G/h-BN with transferred graphene, we employ ARPES with nanometric lateral resolution (NanoARPES\cite{Asensio2, Asensio3, Asensio4}). 
We observe directly renormalisation of the graphene Dirac cone when it crosses the h-BN bands, probably due to the orbital hybridisation between graphene and h-BN, suggesting that orbital character plays a role in shaping the interlayer interaction\cite{Batzill1}.
More importantly, dispersive replicas of h-BN valence band maxima (VBM) are distinguished at the hexagonal Brillouin zone corner. We attribute these replica bands to the formation of Fr\"ohlich polarons, a composite quasiparticle created in this heterostructure by the coupling of phonons and valence electrons. 
In particular, we suggest that the characteristic optical zone-center phonons of $\sim$ 210 meV\cite{Slotman, YSKim} are involved.
To extract physical parameters from this polaronic interaction, we calculate the spectral function with Thomas-Fermi screened coupling using Migdal-Eliashberg theory\cite{YaoWang}, concluding a rather forward scattering with a characteristic distance of 7 unit cells. 
The emergence of the Fr\"ohlich polaron in G/h-BN suggests a practical way to create new composite quasiparticles by atomic layer stacking. Controlling the stacking alignment angle\cite{Novoselov} or interlayer coupling strength\cite{LeRoy3} can potentially tune the properties of the quasiparticle. The potential influences on polaron-related physics are also discussed.

NanoARPES principle, sample preparation and characterisation are detailed in Supplementary Note 1.
Fig. 1a  and b describe the typical NanoARPES setup and sample lattice structure, respectively. The overall photoemission spectra, probed with a beam spot larger than the size of the G/h-BN domain, show sharp core level peaks of carbon 1\textit{s}, boron 1\textit{s} and silicon 2\textit{p} orbitals, suggesting the absence of hydrocarbon contamination (Fig. 1c and Supplementary Note 1).
The crystallographic alignment angle of graphene with respect to h-BN is defined as $\phi$ $\sim$ 3 $^{o}$ as discussed in Supplementary Note 2. 
Fig. 1d shows the real space map of photoemission intensity (NanoARPES image) around the G/h-BN domain. It was collected by integrating the photoemission signal close to the Fermi level while scanning the sample along two in-plane directions with a focused nanometric beam.
This NanoARPES image of the sample matches its optical image with clear contrast (Fig. 1e), demonstrating the capability of our instrument to identify the sample location of both G/h-BN and h-BN samples precisely, with the required lateral resolution and the stability of the whole setup at the sub-micrometric scale.

\begin{figure*}[tbp]
\includegraphics[width=1\textwidth]{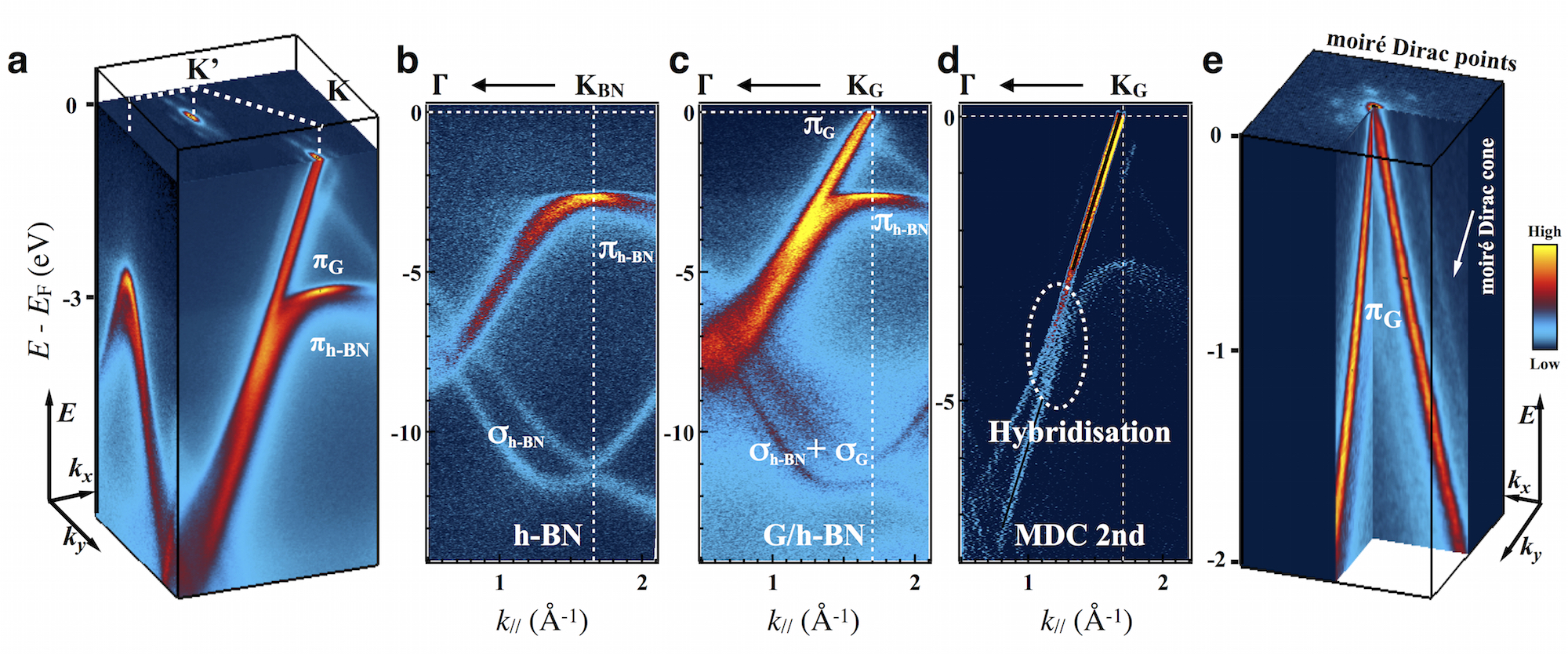}
\caption{\textbf{General electronic structure and Fermi surface of G/h-BN heterostructure and pristine h-BN by NanoARPES}
\textbf{a}, Spectral intensity distribution image in (\textit{k$_{x}$, k$_{y},  E$}) space for G/h-BN. The top surface shows the Fermi surface.
\textbf{b, c}, NanoARPES band structure along $\Gamma$-$\textit{K}$ direction for pristine h-BN and graphene in G/h-BN. 
\textbf{d}, Momentum-distribution-curve-second-derivative (MDC 2nd) spectra along $\Gamma$-$\textit{K}$ direction of graphene in G/h-BN as shown in \textbf{c}. The white dashed circles indicate the band hybridisation region. The black solid lines indicates the Dirac cone dispersion.
\textbf{e}, Detailed Dirac cone spectral intensity distribution of G/h-BN, in (\textit{k$_{x}$, k$_{y},  E$}) space. Again the top surface shows the Fermi surface and emphasizes the existence of moir\'e Dirac cones.
}

\end{figure*}

Taking advantage of the nanometric sample positioning, Fermi surface and valence band structure of G/h-BN and pristine h-BN are collected, as shown in Figure 2.
Although the G/h-BN electronic structure can be naively treated as the superposition of pristine graphene and h-BN valence bands, we observe deviations due to the heterostructure interactions. 
On the one hand, G/h-BN shares similar h-BN $\sigma$- and $\pi$-bands with pristine h-BN. This is reflected in the general consistency of $\pi_{\textrm{{\footnotesize h-BN}}}$ and $\sigma_{{\footnotesize \textrm{h-BN}}}$ between pristine h-BN and G/h-BN (see the comparison in Fig. 2b, c), even with $\sim$ 3 $^{o}$ mismatch. Note that because of the insulating nature of the pristine h-BN, its binding energy is calibrated according to our previous work\cite{Karim}. 
For G/h-BN, electronic states close to the Fermi level are dominated by the monocrystalline graphene $\pi$-band, whose Fermi surface consists of only one main set of Dirac points at the Brillouin zone corners.
On the other hand, the G/h-BN moir\'e superlattice, formed due to the lattice mismatch and rotation, imposes periodic potential on graphene and results in Dirac cone replicas (Fig. 2e).
This observation is consistent with previous works by atomic force microscopy (AFM) and scanning tunneling microscopy (STM)\cite{Geim3, Ashoori, Kim, LeRoy1, Zhang}. 
Analysed in detail in Supplementary Note 2, graphene is rotated $\sim$ 3$^{\circ}$ respect to h-BN and shows no gap opening in the secondary mini Dirac cones\cite{Wang}.
Furthermore, the graphene Dirac cone and h-BN valence band show hybridisation features. This can be intuitively described by the graphene and h-BN bands anti-crossing (see the MDC 2nd spectra in Fig. 2d). 
Similar hybridisation features have also been observed in graphene/MoS$_{2}$ heterostructure\cite{Batzill1}. It is believed that only bands with out-of-plane orbital character are responsible for interlayer interaction and modification of the electronic structure. 
Our findings support this, pointing out the possibility of controlling van der Waals interaction by choosing appropriate materials with different orbital character and twisted angles between them. 

\begin{figure*}[bp]
\centering
\includegraphics[width=1\textwidth]{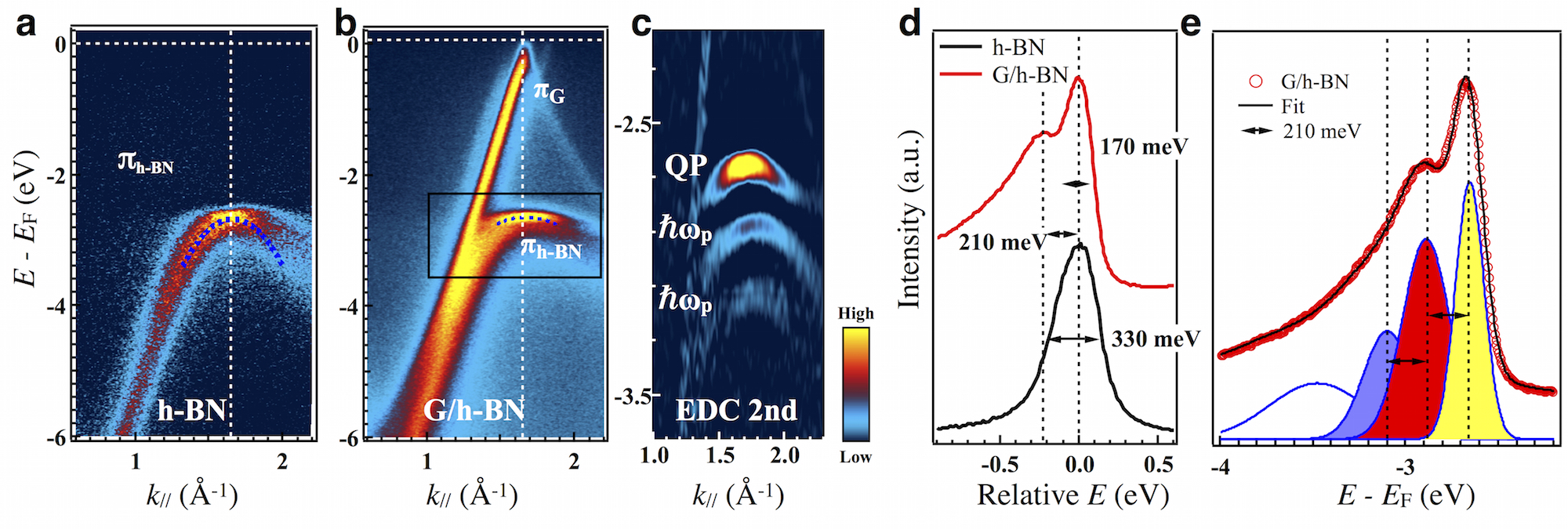}
\caption{\textbf{ Detailed analysis of Fr\"ohlich polaron spectra of G/h-BN}
\textbf{a}, Valence bands of pristine h-BN along $\Gamma$-$\textit{K}$ direction of h-BN.
\textbf{b}, Valence bands of G/h-BN. Blue dashed lines represent the Energy-distribution-curve (EDC)-fitted dispersions. Note that only the $\Gamma$-$\textit{K}$ side is fitted and the $\textit{K}$-$\textit{M}$ side is filled using mirror-plane symmetry. Solid box indicates the region where spectrum in \textbf{c} is from.
\textbf{c}, Zoom in EDC 2nd derivative band structure of G/h-BN.
\textbf{d}, Comparison of VBM spectra from pristine h-BN and graphene covered h-BN. Spectra are shifted in energy to be in line with each other.
\textbf{e}, Poisson fitting to the EDC from the $\textit{K}$ point of G/h-BN. 
}
\end{figure*}

We move forward to analyse in detail the h-BN side in order to examine the exotic phenomena induced by van der Waals interaction.
As shown in Fig. 3a and b, the VBM of pristine h-BN and G/h-BN display significant differences around the $\textit{K}$ point.
For pristine h-BN, the VBM consists of a broad parabolic band, with a fitted effective mass (only for the $\Gamma$-$\textit{K}$ side) $m_{0}$ $\sim$ 0.6 \textit{m$_{e}$} (\textit{m$_{e}$} the free electron mass). In contrast, for G/h-BN the VBM is much flatter with the effective mass fitted as $m^{*}$ $\sim$ 1.4 \textit{m$_{e}$}. 
As analyzed in Supplementary Note 3, single-particle model indicates that coupling to graphene increases the VBM effective mass of h-BN, However, this increase is more than one order of magnitude too small to explain our experimental observations. This suggests that many-body effects are responsible for the renormalization of the effective mass.
Furthermore, if compared with previous ARPES works on h-BN\cite{Karim, Vyalikh}, we observe similar band effective mass and even sharper bands, closer to its intrinsic property. Consequently, it is the construction of heterostructure that induces such effective mass enhancement.

To further understand the quasiparticle physics and interactions, we analyse the energy distribution curves (EDCs) at the VBM. 
As shown in Fig. 3e, NanoARPES spectra display additional peaks at multiples of $\sim$ 210 meV below the G/h-BN VBM. A 2nd derivative analysis (see Fig. 3c) proves the existence of multiple parallel replicas. 
The moir\'e superlattice reciprocal vector (0.16 $\pm$ 0.02 $ \AA^{-1}$) is supposed to duplicate the bands along momentum axis and can not result in such a big energy shift.
In contrast, the equal energy spacing between these replicas suggests energy shake-off due to electron-phonon coupling (EPC)\cite{Shen1, Asensio1, Felix, Grioni, Giustino}. 
In addition, the characteristic energy, $\sim$ 210 meV, coincides with the highest frequency of G/h-BN phonons\cite{Slotman, YSKim}. Note that both graphene and h-BN have phonon frequencies close to this value. Consequently, we cannot identify unambiguously if the phonon involved is from graphene or h-BN. Hence, 
it is more appropriate to associate these replicas to the heterostructure as a whole\cite{Slotman, YSKim}.
Nevertheless, as discussed in the previous paragraph, the band effective mass enhancement only appears in the heterostructure. That means this EPC is activated with the formation of G/h-BN interface. Therefore, this unique electronic structure represents the interfacial coupling of valence electrons (photoholes) and phonons, resulting in the emergence of new composite quasiparticles, 2D Fr\"ohlich polarons. 

To estimate the coupling strength and fit the polaronic spectra for a preliminary understanding, we follow the same procedure as that employed in previous works\cite{Shen1, Asensio1, Felix, Grioni}. The mass enhancement is estimated as $m^{*}/m_{0} \sim 2.3$, suggesting an intermediate EPC. 
Assuming zero-momentum phonons for the EPC, the single-particle spectral function follows a Poisson distribution, as shown in Fig. 3e and Supplementary Note 4. This fitting gives the zero-phonon peak intensity fraction of total weight $\mathit{Z_{0}^{(0)}} \sim 0.34$, corresponding to a 2D Fr\"ohlich coupling constant $\alpha_{2D} \sim 0.9$, according to the diagrammatic quantum Monte Carlo simulation\cite{Svistunov} and scaling relation\cite{Devreese2}.

\begin{figure*}[tbp]
\includegraphics[width=1\textwidth]{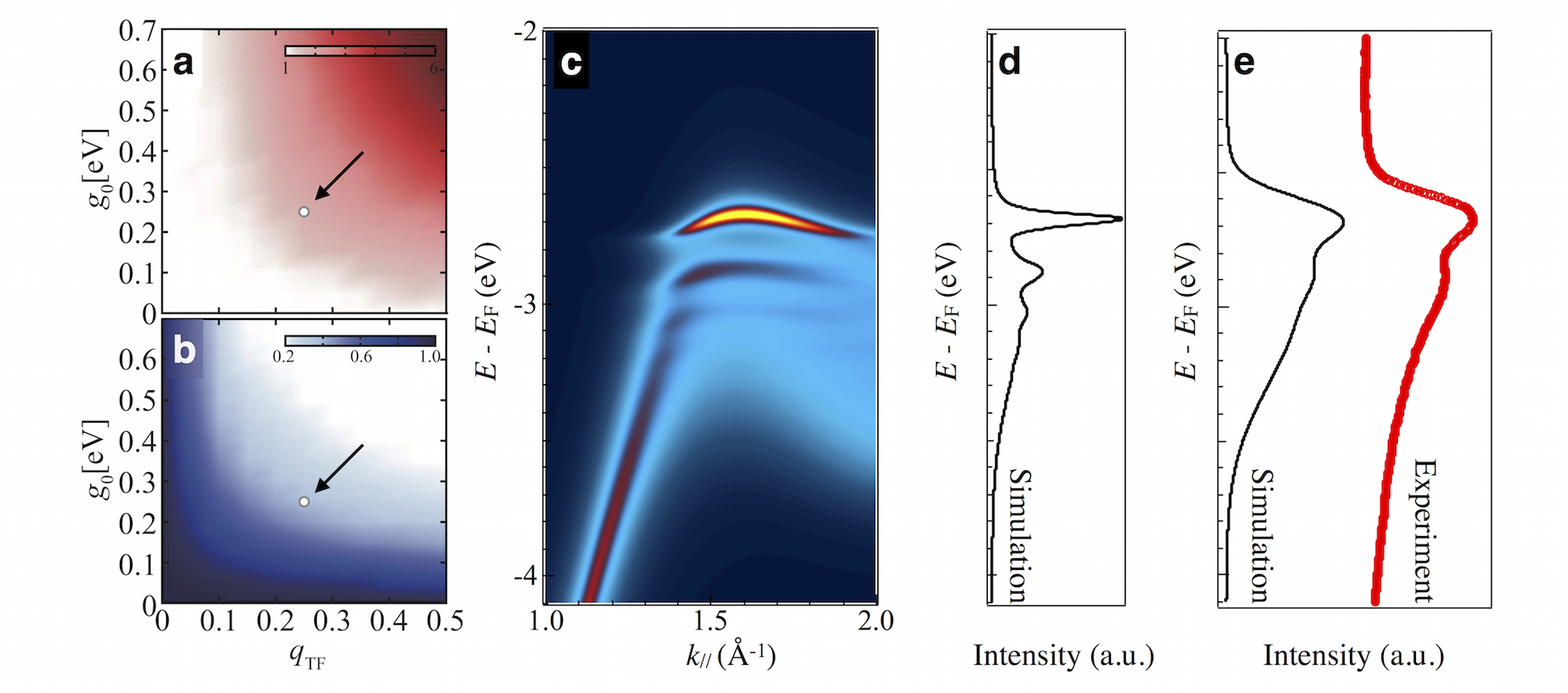}
\caption{\textbf{Simulated polaronic spectra of G/h-BN at 100 K by Migdal-Eliashberg theory}
\textbf{a}, Simulated mass enhancement $m^{*}/m_{0}$ and \textbf{b} quasiparticle residue $\mathit{Z_{0}^{(0)}}$ using various coupling strength $g_0$ and Thomas-Fermi screening wavevector $q_{TF}$ in lattice. The circles indicated by black arrows, denote the optimal parameter ($m^{*}/m_{0} = 2.29$, $\mathit{Z_{0}^{(0)}} = 0.378$) set compared with experiments. 
\textbf{c},  Simulated single-particle spectral function at the optimal parameter set. For simplicity, graphene band is not included.
\textbf{d}, Corresponding simulated EDC though \textit{K} point with Lorentzian FWHM 40 meV.
\textbf{e}, Comparison between simulation and experiment. The simulated EDC is convolved by a Gaussian peak ($\sim$150 meV in width)
}
\end{figure*}

To qualitatively explain the polaron formation and its effective mass enhancement, we calculate the spectral function of G/h-BN with a numerical many-body method by introducing a more realistic, finite momentum, forward scattering.
The valence bands of h-BN is treated by a tight-binding model with up to third nearest neighbors (Supplementary Note 5). The coupling between electrons and phonons is described by the Hamiltonian
\begin{eqnarray}
\Ham = \sum_{\kbf,\sigma} E_{\kbf} c^{\dagger}_{\kbf\sigma}c_{\kbf\sigma} + \sum_{\kbf,\qbf,\sigma} \frac{g(\qbf)}{\sqrt{N}} c^{\dagger}_{\kbf+\qbf,\sigma}c_{\kbf\sigma}\left(a_{\qbf} + a_{-\qbf}^\dagger\right) + \Omega_{0} \sum_{\qbf}  a_{\qbf}^{\dagger} a_{\qbf}
\end{eqnarray}
in which the phonon is simulated by a dispersionless Einstein model with energy $\Omega_0$  and $g(\qbf)$ denotes the electron-phonon coupling vertex. We use Migdal-Eliashberg theory to solve the self-energy and Green's function iteratively (Supplementary Note 5). This theory has been proved suitable to deal with forward scattering for an intermediate coupling\cite{YaoWang}. 

To capture the screen electron-phonon coupling, we choose a Thomas-Fermi coupling vertex as $g(\qbf)=4\pi g_0\left( 1+ |\qbf|^2/q_{\rm TF}^2\right)^2$. Here, $q_{\rm TF}$ presents the Thomas-Fermi screening wavevector, describing the characteristic coupling range in the lattice. In the small $q_{\rm TF}$ limit, this coupling vertex represents the forward scattering, while in large limit a local Holstein coupling is realized.
As shown in Fig. 4a, b, with the increase of $g_0$ and $q_{\rm TF}$, the coupled electron-phonon forms smaller polarons, which is characterised by the effective mass enhancement (Fig. 4a) and quasiparticle residue reduction (Fig. 4b).
The physical $q_{\rm TF}$ and coupling strength $g_0$ are determined through the comparison of experimental effective mass enhancement and the quasiparticle residue $Z_0$. We extract the optimal parameter set $g_0$ = 250 meV and $q_{\rm TF}=0.25$, at which $m^*/m_{0} = 2.29$ and $Z_0 =0.378$. The optimal screening wave vector reflects a characteristic coupling scale of $\sim 7$ unit cells.
Figs. 4c and d show the simulated spectra and EDC from \textit{K} point respectively. To highlight the weak replicas, we use narrow Lorentzian peaks to simulate the h-BN bands. By convolving with a broader Gaussian peak, the simulated EDC shows satisfactory agreement with the experimental EDC, given the fact that for simulation the background is not included (Fig. 4e).

Our findings expand Fr\"ohlich polaron hosts from three-dimensional perovskite surfaces/interfaces\cite{Shen1, Asensio1, Felix, Grioni} to 2D van der Waals heterostructures. It is known that G/h-BN shows enhanced light-matter interaction with unique excitations\cite{Draxl}-polaron could potentially interact with these excitations and give rise to an interesting performance of opto-electronic devices\cite{Novoselov1}.
The large family of 2D materials with diverse band characters allows a flexible approach for exploring polaron physics and its interplay with other emergent phenomena. For example, 2D transition metal dichalcogenides (TMDCs) TiSe$_{2}$ and TiS$_{2}$ have shallow conduction bands. Therefore, it is suitable to construct Fr\"ohlich polarons by transferring graphene on top of these materials. Located close to the Fermi level, these polarons will efficiently modify the macroscopic properties of graphene/TMDC heterostructures, providing an excellent platform for studying the interplay of polarons with other exotic electronic phases such as charge density wave (CDW) and excitonic insulator\cite{Cercellier}.

Our study suggests interfacial coupling as an extra degree of freedom to develop polaron-related physics. Although interfacial polarons have been observed in FeSe/SrTiO$_{3}$ previously\cite{Shen1}, their formation demands a very delicate epitaxial synthesis environment and no parameter is easily accessible to control the polaronic effect. On the contrary, 2D heterostructures can be fabricated with relatively easier controlled parameters such as alignment angle\cite{Geim2, Kim, Geim3, Dean1, LeRoy1, Novoselov, LeRoy3}, component material\cite{Batzill1, Batzill2}, carrier density\cite{Geim2, Kim, Dean1, Asensio3, Batzill2} and so on. This suggests that van der Waals heterostructures are superior to other material systems to serve as a platform for polaron physics study.

\textbf{Acknowledgement}

 The Synchrotron SOLEIL is supported by the Centre National de la Recherche Scientifique (CNRS) and the Commissariat \`a l'Energie Atomique et aux Energies Alternatives (CEA), France. G.Z. thanks the support from the National Basic Research Program of China (973 Program, grant No. 2013CB934500), the National Natural Science Foundation of China (NSFC, grant No. 61325021). The theoretical work was supported at SLAC and Stanford University by the U.S. Department of Energy, Office of Basic Energy Sciences, Division of Materials Sciences and Engineering, under Contract No. DE-AC02-76SF00515. A portion of the computational work was performed using the resources of the National Energy Research Scientific Computing Center supported by the U.S. Department of Energy, Office of Science, under Contract No. DE-AC02-05CH11231. We thank Prof. Shuyun Zhou for inspiring discussions.



\end{document}